\documentclass{article}

\usepackage{graphicx}
\usepackage{a4}

\usepackage[dvips]{color}

\newcommand{\ol}{\overline}

\newcommand{\wt}{\widetilde}

\def\rank{\mathop{\rm rank}\nolimits}

\def\tr{\mathop{\rm tr}\nolimits}

\usepackage[vcentermath]{youngtab}
\newcommand{\Y}{\yng}

\begin{document}
\begin{titlepage}
\title{
\begin{flushright}
\normalsize{ UT-11-03\\
TIT/HEP-608\\
Feb 2011}
\end{flushright}
       \vspace{2cm}
Superconformal index for large $N$ quiver Chern-Simons theories
       \vspace{2cm}}
\author{
Yosuke Imamura\thanks{E-mail: \tt imamura@phys.titech.ac.jp}$^{~1}$,
Daisuke Yokoyama\thanks{E-mail: \tt d.yokoyama@th.phys.titech.ac.jp}$^{~1}$,
and Shuichi Yokoyama\thanks{E-mail: \tt yokoyama@hep-th.phys.s.u-tokyo.ac.jp}$^{~2}$
\\[30pt]
{\it $^1$ Department of Physics, Tokyo Institute of Technology,}\\
{\it Tokyo 152-8551, Japan}\\
{\it $^2$ Department of Physics, University of Tokyo,}\\
{\it Tokyo 113-0033, Japan}
}
\date{}

\maketitle
\thispagestyle{empty}

\vspace{0cm}

\begin{abstract}
\normalsize
We investigate the ${\cal N}=2$ superconformal index for supersymmetric
quiver Chern-Simons theories with large $N$ gauge groups.
After general arguments about the large $N$ limit,
we compute the first few terms in the series expansion
of the index for theories
proposed as dual theories to
homogeneous spaces $V^{5,2}$, $Q^{1,1,1}$, $Q^{2,2,2}$, $M^{1,1,1}$,
and $N^{0,1,0}$.
We confirm that the indices have symmetries expected
from the isometries of dual manifolds.
\end{abstract}

\end{titlepage}


\section{Introduction}
The field-operator correspondence is an important prediction of AdS/CFT\cite{Maldacena:1997re}.
It claims a one-to-one correspondence between gauge invariant operators
in a conformal field theory
and
excitations
in the dual geometry
including
supergravity Kaluza-Klein modes
and extended objects.
It is, in general,  difficult to compute the operator spectrum on
the gauge theory side due to quantum corrections.
However, in supersymmetric theories,
it is often possible to obtain exact results even in the strong coupling region.
For example, in many superconformal field theories, we can exactly compute
superconformal index, which has rich information of the BPS spectrum.
The agreement of the superconformal index on both sides of the duality
provides a strong evidence for the duality.

The ${\cal N}=1$ superconformal index
for four-dimensional gauge theories
is used as a test of AdS/CFT correspondence for ${\cal N}=4$ supersymmetric
Yang-Mills theory in \cite{Kinney:2005ej}.
The agreement of indices on both sides was confirmed to
leading order in the large $N$ limit.
It has been extended to theories with less supersymmetries
\cite{Romelsberger:2005eg,Nakayama:2005mf,
Nakayama:2006ur,Benvenuti:2006qr,Romelsberger:2007ec,
Dolan:2008qi,Spiridonov:2008zr,Gadde:2010en}.

Indices are also applied for analysis of
AdS$_4$/CFT$_3$.
For the ABJM model\cite{Aharony:2008ug}
the ${\cal N}=2$ superconformal index is computed in the perturbative sector\cite{Bhattacharya:2008bja},
which does not contain monopole operators.
It is confirmed that the gauge theory index agrees with that on the gravity side.
This analysis is extended into monopole sectors in \cite{Kim:2009wb},
and the agreement is again confirmed.
Similar analysis is performed for ${\cal N}=3,4,5$ Chern-Simons theories
in \cite{Choi:2008za,Imamura:2009hc,Kim:2010vwa}.

The ${\cal N}=2$ superconformal index is defined by\cite{Bhattacharya:2008bja}
\begin{equation}
I(x,z_i)=\tr\left[(-1)^Fx'^{\{Q,Q^\dagger\}}x^{\Delta+j_3}z_i^{F_i}\right],
\label{idef}
\end{equation}
where the trace is taken over local gauge invariant operators.
$Q$ is a nilpotent supercharge, which is used for the localization.
$\Delta$, $j_3$, and $F_i$ are
the dilatation, the third component of the spin,
and flavor charges.
Our choice of $Q$ is such that it has the quantum numbers
$\Delta=+1/2$, $R=+1$, and $j_3=-1/2$.
 Only BPS states
saturating the BPS bound
\begin{equation}
\{Q,Q^\dagger\}=\Delta-R-j_3\geq0
\label{BPSbound}
\end{equation}
contribute to the index. 
Therefore, it does not depend on $x'$ appearing on the
right hand side in (\ref{idef}).

In the previous works
the canonical conformal dimension of fields are assumed
in computations of the index
by using the localization technique.
This is the case for ${\cal N}\geq3$ Chern-Simons theories
because of non-abelian R-symmetry.
It is recently extended to ${\cal N}=2$ superconformal
field theories with arbitrary R-charge assignments\cite{Imamura:2011su}.
The purpose of this paper is
to rewrite the formula
in \cite{Imamura:2011su} in a form applicable to
large $N$ quiver Chern-Simons theories,
and to compute the index for some examples
of quiver Chern-Simons theories
which are proposed as dual theories to homogeneous $7$-dimensional
Sasaki-Einstein manifolds
(SE$_7$) $V^{5,2}$, $Q^{1,1,1}$, $Q^{2,2,2}$, $M^{1,1,1}$,
and $N^{0,1,0}$.
See \cite{Fabbri:1999hw} and references therein for geometric properties of
these manifolds.

We consider ${\cal N}=2$
superconformal quiver Chern-Simons theories
with the gauge group of the form
\begin{equation}
G=\prod_{A=1}^{n_G}U(N)_A,
\label{gaugegroup}
\end{equation}
and chiral multiplets belonging to bi-fundamental representations.
In \S\ref{flavor.sec} we also introduce flavors,
chiral multiplets belonging to (anti-)fundamental representations.
We denote the Chern-Simons levels by $k_A$.
Namely, the action of the theory contains the Chern-Simons terms
\begin{equation}
\sum_{A=1}^{n_G}\frac{ik_A}{4\pi}\int\tr\left(A_AdA_A-\frac{2i}{3}A_AA_AA_A\right).
\label{csterm}
\end{equation}

In any example of Chern-Simons theory proposed as a dual to
M-theory in a background AdS$_4\times$SE$_7$,
monopole operators play an important role.
For the gauge group (\ref{gaugegroup}),
we can define $n_G$ conserved magnetic charges of an operator by
\begin{equation}
m_A=\frac{1}{2\pi}\oint \tr F_A,
\label{macharge}
\end{equation}
where $F_A$ is the $U(N)_A$ gauge field strength,
and the integration is taken over a sphere
enclosing the insertion point of the operator.
$m_A$ is the monopole charge associated with $U(1)_A$,
the diagonal subgroup of $U(N)_A$.
In theories without flavors,
the magnetic charges are constrained by
Gauss' law constraint
\begin{equation}
\sum_{A=1}^{n_G}k_Am_A=0.
\label{gauss}
\end{equation}

The constraint
(\ref{gauss}) decreases the number of independent magnetic charges by one,
and we have $n_G-1$ independent magnetic charges.
If the theory has the gravity dual,
there should exist corresponding $n_G-1$ charges on the gravity side, too.
In particular, monopole operators with the charge of the form
\begin{equation}
m_1=m_2=\cdots=m_{n_G}=m_{\rm diag}\in{\bf Z},
\label{diagcharge}
\end{equation}
are known to correspond to Kaluza-Klein modes
carrying momentum $m_{\rm diag}$ along the eleventh direction
on the gravity side.
We call such operators diagonal monopole operators.
We need to assume
\begin{equation}
\sum_{A=1}^{n_G}k_A=0,
\end{equation}
for the existence of diagonal monopole operators.
The inclusion of diagonal monopole operators
is essential for the emergence of the eleventh direction.
The supergravity Kaluza-Klein spectrum in AdS$_4\times$SE$_7$
is expected to agree
with the spectrum of operators on the dual conformal field theory
only if we include
both perturbative operators consisting of
elementary fields and diagonal monopole operators.
The purpose of this paper is to obtain non-trivial evidences for
this agreement
by using the ${\cal N}=2$ superconformal index (\ref{idef}).

If $n_G\geq 3$,
we also have non-diagonal monopole operators whose charge is not in the form
(\ref{diagcharge}).
Such non-diagonal monopole operators correspond to M2-branes wrapped on
two-cycles in SE$_7$\cite{Imamura:2008ji}.
When there exist BPS configurations of wrapped M2-branes,
we should take account of them when we compute the index on the gravity side.
This is the case in ${\cal N}=4$ Chern-Simons theories,
whose dual geometries contain vanishing two-cycles\cite{Imamura:2009hc}.
M2-branes wrapped on such shrinking cycles contribute to the
index.
However, in examples we will consider in this paper
all internal spaces
do not have
vanishing two-cycles.
Non-vanishing two-cycles in SE$_7$ are non-BPS\cite{Benishti:2010jn},
and they should decouple in the computation of the index.
Unfortunately, we have not succeeded in proving this decoupling
on the gauge theory side,
and in this paper we simply ignore the contribution of the non-diagonal monopole
operators.
In flavored theories,
there may exist operators whose magnetic charges
do not satisfy the constraint
(\ref{gauss}).
We also ignore such operators, and we focus only on
operators with diagonal magnetic charges.

We rely on a numerical method to obtain the index as the series expansion
with respect to $x$.
We compute the index in all examples up to the order of $x^2$.
This is mainly because it takes much longer time to obtain higher order terms
than ${\cal O}(x^2)$.

This paper is organized as follows.
In the next section, we write down the formula derived
in \cite{Imamura:2011su} in the case of quiver Chern-Simons theories
without flavors.
In \S\ref{largen.sec} and \S\ref{pref.sec} we take the large $N$ limit and derive
a formula for the index for vector-like theories.
We apply it to a theory proposed as a dual theory to $V^{5,2}$ in \S\ref{v52.sec},
and confirm that the index has the symmetry which is
expected from the
isometry of $V^{5,2}$.
In \S\ref{chiral.sec}
we comment on the factorization of the index for chiral theories.
We compute index for theories dual to $Q^{1,1,1}$,
$Q^{2,2,2}$, and $M^{1,1,1}$ in
\S\ref{q111.sec},
\S\ref{q222.sec}, and
\S\ref{m111.sec}, respectively.
We again confirm that in all cases the index has
desirable symmetry.
In \S\ref{flavor.sec} we extend the formula to theories
with flavors, and
in \S\ref{n010.sec} we apply it to a
theory dual to $N^{0,1,0}$.
The last section summarizes the results.

\section{Formula for the index}
The index (\ref{idef}) can be defined as the path integral of the
theory in the background ${\bf S}^1\times{\bf S}^2$
with appropriate boundary conditions.
A formula for the index (\ref{idef}) is derived from
this path integral with the help of localization method
associated with the supercharge $Q$\cite{Kim:2009wb,Imamura:2011su}.
If we deform the action by appropriate $Q$-exact terms,
the path integral is localized at saddle points
corresponding to GNO monopoles\cite{Goddard:1976qe},
which are Dirac monopoles for the Cartan part of the gauge group $G$.
The magnetic charge of a GNO monopole is specified by
a set of $\rank G=n_GN$ integers, defined by
\begin{equation}
m_{A,i}=\frac{1}{2\pi}\oint_{{\bf S}^2} F_{A,i},
\label{maidef}
\end{equation}
where $F_{A,i}$ is the $i$-th diagonal
component of the  $U(N)_A$ gauge flux $F_A$.
The conserved charges defined in (\ref{macharge})
are related to $m_{A,i}$ by
\begin{equation}
m_A=\sum_{i=1}^Nm_{A,i}.
\end{equation}
Note that $m_{A,i}$ are not conserved quantities,
unlike $m_A$.
We should regard them as integers labeling
saddle points dominating the path integral.
In addition to the monopole charges $m_{A,i}$,
saddle points are also labeled by holonomy, denoted by
$a$, which is the Wilson line around the ${\bf S}^1$.
Namely, the saddle points form continuous sets in the configuration
space.
$m$ and $a$ take values in the Cartan part of the Lie algebra
of the gauge group $G$.

We need to sum the contribution over all the saddle points.
Let us denote this summation in a symbolic way by
\begin{equation}
\sum_m \int da.
\end{equation}
$\sum_m$ is the summation over all independent magnetic charges.
Here $m$ means a set of $n_GN$ magnetic charges $m_{A,i}$.
We regard two magnetic charges transformed to each other by the Weyl Group of $G$
as being equivalent.
By using this equivalence,
we arrange the components of monopole charges
in descending order in the set of $N$ components
for each $U(N)$ gauge group.
$\int da$ is given by
\begin{equation}
\int da
=\frac{1}{(\mbox{stat})}\prod_{A=1}^{n_G}\prod_{\rho\in{\bf N}_A}
\int \frac{d\rho(a)}{2\pi}
=\frac{1}{(\mbox{stat})}\prod_{A=1}^{n_G}\prod_{i=1}^N
\int \frac{da_{A,i}}{2\pi}.
\label{measure}
\end{equation}
We pick up a component of $a$ corresponding to
a $U(1)$ subgroup of $U(N)_A$
by applying a weight $\rho\in{\bf N}_A$,
where ${\bf N}_A$ is the fundamental representation
of $U(N)_A$.
$(\mbox{stat})$ is a numerical factor
defined in the following way.
Let us focus on one of $U(N)$ factors.
In general, it is broken by the magnetic flux $m$ to
its subgroup in the form
\begin{equation}
\prod_i U(N_i)
\subset U(N).
\end{equation}
The order of the Weyl group of this
unbroken symmetry is $\prod_iN_i!$.
The statistical factor appearing in
(\ref{measure}) is the product of this number for
all $U(N)$ factors.

The trace in the definition of the superconformal index (\ref{idef})
represents the summation over all multi-particle states in ${\bf S}^2$,
including single-particle states and the vacuum state.
To compute the index (\ref{idef}), it is convenient to define the
letter index
\begin{equation}
f(e^{ia},x,z_i)
=\sum(-1)^F e^{i\rho(a)}x'^{\Delta-R-j_3}x^{\Delta+j_3}z_i^{F_i}.
\label{defletter}
\end{equation}
The letter index is defined for each monopole background, and
the summation in (\ref{defletter}) is taken over
all single particle states on a GNO monopole background.
The states summed over in (\ref{defletter})
in general have non-vanishing gauge charges.
We use the Wilson line $a$ as a chemical potential for
the gauge charges.
Once we obtain the letter index,
the index for excitations including arbitrary number of particles
is given as the plethystic exponential of the letter index
\begin{equation}
\exp\left[\sum_{n=1}^\infty\frac{1}{n}f(e^{ina},x^n,z_i^n)\right].
\label{excitation}
\end{equation}

We also need to take account of the quantum numbers of the vacuum state,
which are obtained by summing up the zero-point contribution of
all single-particle states summed in (\ref{defletter}).
(Usually they diverge and we need suitable regularizations.)
This is expressed in the form
\begin{equation}
P=e^{-S_{CS}}e^{ib_0(a)}x^{\epsilon_0}z_i^{q_{0i}}.
\label{prefactor}
\end{equation}
We call this a prefactor.
$x^{\epsilon_0}$, $z_i^{q_{0i}}$, and $e^{ib_0(a)}$ correspond to
the zero-point energy, zero-point flavor charges, and zero-point
gauge charges, respectively, carried by the vacuum state.
$S_{\rm CS}$
is the classical contribution
from Chern-Simons terms (\ref{csterm}),
\begin{equation}
S_{\rm CS}
=i\sum_{A=1}^{n_G}\sum_{\rho\in{\bf N}_A}k_A\rho(a)\rho(m),
\label{scs0}
\end{equation}
and has the effect of shifting gauge charges of the vacuum state.

We obtain the complete superconformal index
by summing up the contribution of all saddle points\cite{Kim:2009wb,Imamura:2011su},
\begin{equation}
I(x,z_i)
=\sum_m\int da P
\exp\left[\sum_{n=1}^\infty\frac{1}{n}f(e^{ina},x^n,z_i^n)\right].
\label{formula}
\end{equation}
The integral over the continuous parameter $a$ picks up
only the contribution of gauge invariant states.

The explicit form of the letter index
$f$ is given in \cite{Kim:2009wb,Imamura:2011su}.
It is the sum of the vector multiplet contribution
$f_{\rm vector}$ and the chiral multiplet contribution
$f_{\rm chiral}$.
For a quiver gauge theory with gauge group (\ref{gaugegroup})
and bi-fundamental chiral multiplets,
they are given by
\begin{eqnarray}
&&f_{\rm vector}(e^{ia},x)
=\sum_{A=1}^{n_G}
\sum_{\rho\in{\bf N}_A}
\sum_{\rho'\in{\bf N}_A}
(1-\delta_{\rho,\rho'})
\left(-e^{i(\rho(a)-\rho'(a))}x^{|\rho(m)-\rho'(m)|}\right),
\\
&&f_{\rm chiral}(e^{ia},x,z_i)
=
\sum_{\Phi_{AB}}
\sum_{\rho\in{\bf N}_A}
\sum_{\rho'\in{\bf N}_B}
\frac{x^{|\rho(m)-\rho'(m)|}}{1-x^2}
\nonumber\\&&\hspace{5em}
\left(
e^{i(\rho(a)-\rho'(a))}z_i^{F_i(\Phi)}x^{\Delta(\Phi)}
-e^{-i(\rho(a)-\rho'(a))}z_i^{-F_i(\Phi)}x^{2-\Delta(\Phi)}
\right).
\label{fchdef}
\end{eqnarray}
$\sum_{\Phi_{AB}}$ is the summation over all bi-fundamental fields.
To indicate that a chiral multiplet $\Phi$ belongs to
$({\bf N}_A,\ol{\bf N}_B)$, we use the notation
$\Phi_{AB}$.

With an appropriate regularization,
we obtain the zero-point contributions\cite{Kim:2009wb,Imamura:2011su}
\begin{eqnarray}
\epsilon_0
&=&\frac{1}{2}\sum_{\Phi_{AB}}
\sum_{\rho\in{\bf N}_A}\sum_{\rho'\in{\bf N}_B}|\rho(m)-\rho'(m)|(1-\Delta(\Phi))
\nonumber\\&&\hspace{5em}
-\frac{1}{2}\sum_{A=1}^{n_G}
\sum_{\rho\in{\bf N}_A}\sum_{\rho'\in{\bf N}_A}|\rho(m)-\rho'(m)|,
\label{ep0}\\
q_{0i}
&=&-\frac{1}{2}\sum_{\Phi_{AB}}
\sum_{\rho\in{\bf N}_A}\sum_{\rho'\in{\bf N}_B}|\rho(m)-\rho'(m)|F_i(\Phi),
\label{q0i}
\\
b_0(a)
&=&-\frac{1}{2}\sum_{\Phi_{AB}}
\sum_{\rho\in{\bf N}_A}\sum_{\rho'\in{\bf N}_B}|\rho(m)-\rho'(m)|
(\rho(a)-\rho'(a)).
\label{b0}
\end{eqnarray}
For vector-like theories, $b_0(a)$ identically vanishes.

\section{Large $N$ limit and the factorization}\label{largen.sec}
The prediction of field-operator correspondence in the large $N$ limit
is expressed as
\begin{equation}
I(x,z_i)=\exp\left(\sum_{n=1}^\infty\frac{1}{n}I^{\rm sp}(x^n,z_i^n)\right),
\label{mpsp}
\end{equation}
where $I(x,z_i)$ is the superconformal index (\ref{formula}) given in the last section
and 
$I^{\rm sp}$ is a supergravity single-particle index,
which represents the spectrum of one-particle states on the gravity side.

In the large $N$ limit,
the formula (\ref{formula}) includes
an infinite number of integrals.
These integrals are treated in the following way.
We first factorize 
the index into three parts\cite{Kim:2009wb,Imamura:2011su},
\begin{equation}
I=I^{(0)}I^{(+)}I^{(-)},\quad
I^{(\pm)}=\sum_{m^{(\pm)}}I^{(\pm)}_{m^{(\pm)}}.
\label{factorize}
\end{equation}
The factors $I^{(+)}$ and $I^{(-)}$ depend only on magnetic charges
$m^{(+)}$ and $m^{(-)}$, respectively, while the other factor $I^{(0)}$
is independent of magnetic charges.
$m^{(\pm)}$ are positive and negative
part of $m$.
For example, if the gauge group is $G=U(5)^3$ and $m$ has the components
\begin{equation}
m=(2,1,0,-2,-3;1,0,0,0,0;3,0,0,0,-1),
\label{mexp1}
\end{equation}
then the positive and negative parts are
\begin{equation}
m^{(+)}=(2,1;1;3),\quad
m^{(-)}=(-2,-3;;-1).
\end{equation}
We represent these with Young diagrams as
\begin{equation}
m^{(+)}=(\Y(2,1),\Y(1),\Y(3)),\quad
m^{(-)}=-(\Y(3,2),\cdot,\Y(1)).
\end{equation}
Because we focus only on diagonal monopole operators
as we mentioned in Introduction,
the summation with respect to $m^{(+)}$ in (\ref{factorize})
is taken over all monopole charges satisfying
(\ref{diagcharge}).
For example,
in the case of gauge group $U(N)^2$
and Chern-Simons levels $(k,-k)$,
it is
\begin{equation}
I^{(+)}
=I^{(+)}_{(\cdot,\cdot)}
+I^{(+)}_{(\Y(1),\Y(1))}
+I^{(+)}_{(\Y(2),\Y(2))}
+I^{(+)}_{(\Y(2),\Y(1,1))}
+I^{(+)}_{(\Y(1,1),\Y(2))}
+I^{(+)}_{(\Y(1,1),\Y(1,1))}
+\cdots.
\end{equation}
The first term $I^{(+)}_{(\cdot,\cdot)}$ is
always $1$.
$I^{(-)}$ is also given as a similar series.
Once we obtain a formula in the form
(\ref{factorize}), we can calculate the index
numerically with computers as an $x$ expansion.
The aim of this and the next section is to rewrite the formula
given in the last section into this factorized form
in the case of large $N$ quiver Chern-Simons theories.

Corresponding to the decomposition of $m$ into $m^{(\pm)}$
(and the remaining part consisting of vanishing components),
we decompose $a$ into three parts.
For example, if $m$ is given by (\ref{mexp1}),
the three parts of $a=(a_{1,1},a_{1,2},\ldots,a_{3,5})$ are
\begin{eqnarray}
a^{(+)}&=&(a_{1,1},a_{1,2};a_{2,1};a_{3,1}),\\
a^{(-)}&=&(a_{1,4},a_{1,5};;a_{3,5}),\\
a^{(0)}&=&(a_{1,3};a_{2,2},a_{2,3},a_{2,4},a_{2,5};a_{3,2},a_{3,3},a_{3,4}).
\end{eqnarray}
Namely, if $m_{A,i}$ is zero (positive, negative),
the corresponding element $a_{A,i}$ belongs to $a^{(0)}$
($a^{(+)}$, $a^{(-)}$).
In the large $N$ limit, we keep the number of
non-vanishing components of the magnetic flux
to be order $1$, while the number of components of
$a^{(0)}$ grows as ${\cal O}(N)$.
It is necessary to
perform the integration over $a^{(0)}$ analytically,
before we rely on the numerical computation.
We follow the prescription given in \cite{Kim:2009wb}.
We first decompose the letter index into two parts.
Let us first consider $f_{\rm chiral}$.
It takes the form
\begin{equation}
f_{\rm chiral}(e^{ia},x,z_i)=\sum x^{|\rho(m)-\rho'(m)|}(\cdots).
\end{equation}
We define $f'_{\rm chiral}$ by replacing $|\rho(m)-\rho'(m)|$
in $f_{\rm chiral}$
by $|\rho(m)|+|\rho'(m)|$, and denote the remaining part
by $f_{\rm chiral}^{\rm mod}$,
\begin{eqnarray}
f'_{\rm chiral}(e^{ia},x,z_i)
&=&
\sum x^{|\rho(m)|+|\rho'(m)|}(\cdots),\\
f^{\rm mod}_{\rm chiral}(e^{ia},x,z_i)
&=&\sum
(x^{|\rho(m)-\rho'(m)|}-x^{|\rho(m)|+|\rho'(m)|})
(\cdots).
\end{eqnarray}
An important property of
$f^{\rm mod}_{\rm chiral}$ is
that it vanishes unless
$\rho(m)$ and $\rho'(m)$ are both positive or both negative.
Thanks to this property, we can further decompose
$f_{\rm chiral}^{\rm mod}$ into positive part $f^{(+)}_{\rm chiral}$
and negative part $f^{(-)}_{\rm chiral}$
defined by
\begin{equation}
f^{(\pm)}_{\rm chiral}(e^{ia},x,z_i)
=
\sum_{\Phi_{AB}}
\sum_{\rho\in{\bf N}_A}^{(\pm)}
\sum_{\rho'\in{\bf N}_B}^{(\pm)}
(x^{|\rho(m)-\rho'(m)|}-x^{|\rho(m)|+|\rho'(m)|})
(\cdots),
\end{equation}
where $\sum^{(+)}_{\rho\in{\bf N}_A}$ and $\sum^{(-)}_{\rho\in{\bf N}_A}$
represent the summation over $\rho\in{\bf N}_A$ satisfying $\rho(m)>0$
and  $\rho(m)<0$, respectively.

Similarly, we define $f'_{\rm vector}$ and $f_{\rm vector}^{(\pm)}$ by
\begin{eqnarray}
&&f'_{\rm vector}(e^{ia},x)
=-\sum_{A=1}^{n_G}
\sum_{\rho\in{\bf N}_A}
\sum_{\rho'\in{\bf N}_A}
x^{|\rho(m)|+|\rho'(m)|}e^{i(\rho(a)-\rho'(a))},\\
&&f^{(\pm)}_{\rm vector}(e^{ia},x)
=\sum_{A=1}^{n_G}
\sum_{\rho\in{\bf N}_A}^{(\pm)}
\sum_{\rho'\in{\bf N}_A}^{(\pm)}
\nonumber\\&&\hspace{3em}
\Big[-(1-\delta_{\rho,\rho'})
x^{|\rho(m)-\rho'(m)|}
+x^{|\rho(m)|+|\rho'(m)|}
\Big]
e^{i(\rho(a)-\rho'(a))}.
\end{eqnarray}

$f^{(\pm)}=f_{\rm chiral}^{(\pm)}+f_{\rm vector}^{(\pm)}$ do not depend
on $a^{(0)}$.
If we assume that the theory is vector-like,
the prefactor does not depend on $a^{(0)}$, either.
Components of $a^{(0)}$ appear in the integrand of
(\ref{formula}) only through $f'=f'_{\rm chiral}+f'_{\rm vector}$.
If we define $\lambda_{A,n}$ by
\begin{equation}
\lambda_{A,n}
=
\sum_{\rho\in {\bf N}_A}
x^{|n\rho(m)|}
e^{in\rho(a)},
\label{lambdadef}
\end{equation}
we can rewrite $f'$ in the quadratic form of $\lambda_{A,n}$,
\begin{eqnarray}
f'(e^{ia},x,z_i)
&=&-\sum_{A=1}^{n_G}\lambda_{A,+1}\lambda_{A,-1}
\nonumber\\
&&
+\sum_{\Phi_{A,B}}
\left[
\lambda_{A,+1}
\lambda_{B,-1}
\frac{z_i^{F_i}x^{\Delta(\Phi)}}{1-x^2}
-
\lambda_{A,-1}
\lambda_{B,+1}
\frac{z_i^{-F_i}x^{2-\Delta(\Phi)}}{1-x^2}
\right]
\nonumber\\
&=&
-\sum_{A,B}\lambda_{A,+1}M_{A,B}(x,z_i)\lambda_{B,-1}.
\label{fpc}
\end{eqnarray}
$M_{A,B}$ is the matrix whose
components are read off from (\ref{fpc}).

The exponential factor in the integrand in (\ref{formula})
is now factorized into three parts,
\begin{equation}
\exp\left(\sum_{n=1}^\infty\frac{1}{n}f^{(+)}(\cdot^n)\right)
\exp\left(\sum_{n=1}^\infty\frac{1}{n}f^{(-)}(\cdot^n)\right)
\exp\left(\sum_{n=1}^\infty\frac{1}{n}f'(\cdot^n)\right).
\label{exp3}
\end{equation}
We use $(\cdot^n)$ for the arguments
which are replaced by $n$-th power of original ones.
In (\ref{exp3}) it represents $(e^{ina},x^n,z_i^n)$.
Only the last factor contains $a^{(0)}$.
Because it depends on $a^{(0)}$ only through $\lambda_{A,n}$,
we can change the integration variables
from $a^{(0)}$ to $\lambda_{A,n}$.
It is known that the Jacobian factor associated with this variable change is constant
if the integral is dominated by the uniform
eigenvalue distribution of $a^{(0)}$.
Although the possibility of phase transition to the deconfined phase
in a certain example of AdS$_3$/CFT$_4$ is pointed out in \cite{Kim:2010vwa},
we assume the domination of
the uniform eigenvalue distribution.
Then $\lambda_{A,n}$ integrals become the
Gaussian integral,
\begin{eqnarray}
I^{(0)}(x,z_i)
&=&\int d\lambda\exp\left(-\sum_{n=1}^\infty\frac{1}{n}\sum_{A,B}M_{A,B}(x^n,z_i^n)
\lambda_{A,n}\lambda_{B,-n}
\right)
\nonumber\\
&=&\frac{1}{\prod_{n=1}^\infty\det M_{A,B}(x^n,z_i^n)}.
\label{neutralint2}
\end{eqnarray}
Assuming the factorization of the prefactor $P=P^{(+)}P^{(-)}$,
we factorize the index into three parts as (\ref{factorize}),
and $I^{(\pm)}$ are given by
\begin{equation}
I^{(\pm)}(x,z_i)=
\sum_{m^{(\pm)}}
\int da^{(\pm)}
P^{(\pm)}(e^{ia},x,z_i)
\exp\left(\sum_{n=1}^\infty\frac{1}{n}f^{(\pm)}(e^{ina},x^n,z_i^n)\right).
\end{equation}
For each $m^{(\pm)}$, this contains a finite number of integrals.

The remaining task is to show the factorization of the prefactor $P$.
\section{Factorization of the prefactor}\label{pref.sec}
Let us prove the factorization of the prefactor $P$ in (\ref{prefactor}).
Here we assume that the theory is vector-like and $b_0(a)$ identically vanishes.
The prefactor then consists of three factors,
\begin{equation}
P=
e^{-S_{\rm CS}}
x^{\epsilon_0}z_i^{q_{0i}}.
\end{equation}
We easily see the factorization of the first factor,
\begin{equation}
e^{-S_{\rm CS}}
=e^{-S_{\rm CS}^{(+)}}e^{-S_{\rm CS}^{(-)}},\quad
S_{\rm CS}^{(\pm)}
=i\sum_{A=1}^{n_G}\sum_{\rho\in{\bf N}_A}^{(\pm)}k_A\rho(a)\rho(m).
\end{equation}

To show the factorization of $x^{\epsilon_0}$ and $z_i^{q_{0i}}$,
we follow the prescription we used for the letter index.
We define $\epsilon_0'$ and $q'_{0i}$
by replacing
the factor $|\rho(m)-\rho'(m)|$ in $\epsilon_0$ and $q_{0i}$ by
$|\rho(m)|+|\rho'(m)|$,
and $\epsilon_0^{\rm mod}$
and $q_{0i}^{\rm mod}$ as the remaining parts.
$\epsilon_0^{\rm mod}$ and $q_{0i}^{\rm mod}$
contain the factor
\begin{equation}
|\rho(m)-\rho'(m)|-|\rho(m)|-|\rho'(m)|,
\end{equation}
and this is non-vanishing only when
$\rho(m)$ and $\rho'(m)$ are both positive or both negative.
We can decompose $\epsilon_0^{\rm mod}$ and $q_{0i}^{\rm mod}$
into positive and negative parts,
\begin{eqnarray}
\epsilon_0^{(\pm)}
&=&\frac{1}{2}\sum_{\Phi_{AB}}
\sum_{\rho\in{\bf N}_A}^{(\pm)}
\sum_{\rho'\in{\bf N}_B}^{(\pm)}
(|\rho(m)-\rho'(m)|-|\rho(m)|-|\rho'(m)|)(1-\Delta(\Phi))
\nonumber\\&&\hspace{3em}
-\frac{1}{2}\sum_{A=1}^{n_G}
\sum_{\rho\in{\bf N}_A}^{(\pm)}\sum_{\rho'\in{\bf N}_A}^{(\pm)}
(|\rho(m)-\rho'(m)|-|\rho(m)|-|\rho'(m)|),
\label{ep0pm}\\
q_{0i}^{(\pm)}
&=&-\frac{1}{2}\sum_{\Phi_{AB}}
\sum_{\rho\in{\bf N}_A}^{(\pm)}
\sum_{\rho'\in{\bf N}_A}^{(\pm)}
(|\rho(m)-\rho'(m)|-|\rho(m)|-|\rho'(m)|)F_i(\Phi).
\label{q0ipm}
\end{eqnarray}

We rewrite $\epsilon_0'$ and $q_{0i}'$ as
\begin{eqnarray}
\epsilon'_0
&=&N\sum_{A=1}^{n_G} \beta_A\sum_{\rho\in{\bf N}_A}|\rho(m)|,
\label{eps0p}\\
q'_{0i}&=&-N\sum_{A=1}^{n_G}\eta_{A,i}
\sum_{\rho\in{\bf N}_A}|\rho(m)|.
\label{qp0i}
\end{eqnarray}
where $\beta_A$ and $\eta_{A,i}$ are defined by
\begin{eqnarray}
\beta_A&=&\frac{1}{2}\sum_{\Phi\in A}(1-\Delta(\Phi))-1,
\label{betaa}\\
\eta_{A,i}&=&\frac{1}{2}\sum_{\Phi\in A}F_i(\Phi).
\label{fa}
\end{eqnarray}
$\sum_{\Phi\in A}$ represents
summation over bi-fundamental fields
coupled by the $U(N)_A$ gauge field.
If there are $U(N)_A$ adjoint chiral multiplets,
they should be taken twice.
It is obvious that we can divide
(\ref{eps0p}) and (\ref{qp0i}) into positive and negative
parts depending only on $m^{(+)}$ and $m^{(-)}$,
respectively.

In the four-dimensional ${\cal N}=1$ supersymmetric
quiver gauge theory
described by the same quiver diagram,
$\beta_A$ and $\eta_{A,i}$ are the coefficients of the NSVZ exact
$\beta$-functions\cite{Novikov:1983uc,Novikov:1985ic,Novikov:1985rd} of
$SU(N)_A$ gauge groups
and the coefficients of the $\tr U(1)_{F_i}SU(2)_A^2$ anomaly,
respectively.
These quantities should vanish if the four-dimensional
theory is conformal and the flavor-symmetries are
anomaly free.
Of course they do not have to vanish in three dimensional theories.
It is not clear why the exact $\beta$-functions and the anomaly coefficients
appear here.
It may be interesting to consider physical
implication of the appearance of such quantities in three-dimensional
gauge theories.

Although we have shown the factorization of
$x^{\epsilon_0'}$ and $z_i^{q_{0i'}}$,
we do not have to take them into account
in the following
because
in all examples below $\epsilon_0'$ and $q_{0i}'$ vanish for
diagonal monopole operators.
In this case
$I^{(+)}$ is given by
\begin{equation}
I^{(+)}(x,z_i)
=
\sum_{m^{(+)}}\int da^{(+)}
e^{-S_{\rm CS}^{(+)}}x^{\epsilon_0^{(+)}}z_i^{q_{0i}^{(+)}}
\exp\left[\sum_{n=1}^\infty\frac{1}{n}
f^{(+)}(e^{ina},x^n,z_i^n)
\right].
\label{formulaplus}
\end{equation}
The formula for $I^{(-)}$ is obtained by replacing all $+$ by $-$
in (\ref{formulaplus}).
$I^{(+)}$ and $I^{(-)}$ are related by
the charge conjugation,
which reverses the orientation of arrows in the quiver diagram.
If the reversal of arrows does not change the theory,
we can immediately obtain $I^{(-)}$ from $I^{(+)}$.
See the following examples for concrete relations between $I^{(+)}$ and $I^{(-)}$.

\section{Example 1: $V^{5,2}$}\label{v52.sec}
\begin{figure}[htb]
\centerline{\includegraphics{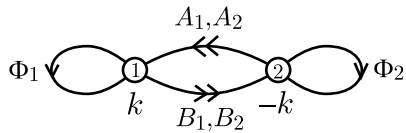}}
\caption{The quiver diagram of a Chern-Simons theory dual to $V^{5,2}/{\bf Z}_k$.}
\label{v52quiv}
\end{figure}

$V^{5,2}$ is the homogeneous space defined as the coset
\begin{equation}
V^{5,2}=SO(5)/SO(3),
\end{equation}
and has the isometry
\begin{equation}
SO(5)\times SO(2).
\end{equation}
The cone over this manifold is the non-compact Calabi-Yau
$4$-fold
\begin{equation}
v_1^2+v_2^2+v_3^2+v_4^2+v_5^2=0.
\label{v52cone}
\end{equation}
The $SO(5)$ isometry is the rotations
mixing five variables $v_i$,
while $SO(2)$ is the simultaneous phase rotation of $v_i$.
A dual Chern-Simons theory for $V^{5,2}$ is proposed in \cite{Martelli:2009ga}.
It is the vector-like quiver gauge theory shown in Fig.\ \ref{v52quiv}.
The superpotential is
\begin{equation}
W=\tr(\Phi_1^3-\epsilon^{ij}\Phi_1A_iB_j+\epsilon^{ji}B_jA_i\Phi_2-\Phi_2^3).
\end{equation}
The manifest global symmetry of this theory is
\begin{equation}
SU(2)\times U(1)_B\times U(1)_R,
\label{v52cssym}
\end{equation}
where $SU(2)$ rotates $A_i$ and $B_i$ simultaneously as doublets,
and $U(1)_B$ is the baryonic symmetry.%
\footnote{The baryonic symmetry in a quiver Chern-Simons theory
is the symmetry generated by $B\propto\sum_{A=1}^{n_G}k_AT_A$ where
$T_A$ are the generators of $U(1)_A$.
We include the baryonic symmetry in flavor symmetries.}
Let $F_1$ and $F_2$ be the generators of the $SU(2)$ Cartan
and $U(1)_B$, respectively.
The charge assignments of these flavor symmetries and
the R-charge assignment determined by assuming the marginality
of the terms in the superpotential are shown in Table \ref{table:v52}.
\begin{table}[htb]
\caption{Charge assignments of R and flavor symmetries
for the theory in Fig.\ \ref{v52quiv} are shown.}
\label{table:v52}
\begin{center}
\begin{tabular}{ccccccc}
\hline
\hline
      & $\Phi_1$ & $\Phi_2$  & $A_1$ & $A_2$ & $B_1$ & $B_2$ \\
\hline
$\Delta$ & $\frac{2}{3}$ & $\frac{2}{3}$ & $\frac{2}{3}$ & $\frac{2}{3}$ & $\frac{2}{3}$ & $\frac{2}{3}$ \\
$F_1$ & $0$ & $0$ & $\frac{1}{2}$ & $-\frac{1}{2}$ & $\frac{1}{2}$ & $-\frac{1}{2}$ \\
$F_2$ & $0$ & $0$ & $\frac{1}{2}$ & $\frac{1}{2}$ & $-\frac{1}{2}$ & $-\frac{1}{2}$ \\
\hline
\end{tabular}
\end{center}
\end{table}
For general $k$, this theory is dual to $V^{5,2}/{\bf Z}_k$ where ${\bf Z}_k$
is generated by the rotation of two-dimensional complex vectors $(v_2,v_3)$ and $(v_4,v_5)$
by angle $2\pi/k$.
For $k=2$, the $SO(5)$ factor of the isometry is broken to
$O(4)$,
and for $k\geq3$ to $SU(2)\times U(1)$.

Due to the manifest $SU(2)$ symmetry, the index satisfies
the relation
\begin{equation}
I(x,z_1,z_2)
=I(x,z_1^{-1},z_2),
\label{v52sym1}
\end{equation}
corresponding to the Weyl group of $SU(2)$.

If we apply the charge conjugation to this theory by reversing
the direction of all arrows in the quiver diagram,
we obtain the theory with $A_i$ and $B_i$
exchanged.
This implies the relation between $I^{(\pm)}$,
\begin{equation}
I^{(-)}(x,z_1,z_2)
=I^{(+)}(x,z_1,z_2^{-1}),
\label{v52symc}
\end{equation}
and the complete index satisfies
\begin{equation}
I(x,z_1,z_2)
=I(x,z_1,z_2^{-1}).
\label{v52sym2}
\end{equation}

In the case of $k=1$, the dual geometry is $V^{5,2}$, and
the flavor symmetry $SU(2)\times U(1)_B$ is expected to
be enhanced to $SO(5)$.
If so, the index should be invariant under the Weyl group
of $SO(5)$.
This means that the index should satisfy
\begin{equation}
I(x,z_1,z_2)=I(x,z_2,z_1)
\end{equation}
in addition to (\ref{v52sym1}) and (\ref{v52sym2}).
Let us confirm that this is actually the case by computing the index.

The neutral part of the index up to the order of $x^2$ is
\begin{equation}
I^{(0)}=1+2x^{2/3}+(\chi_1(z_1)+4)x^{4/3}+(6+2\chi_1(z_1))x^{2}+\cdots,
\label{v52i0}
\end{equation}
where
$\chi_s$ is the $SU(2)$ character,
\begin{equation}
\chi_s(z)=\frac{z^{s+1}-z^{-s}}{z-1}=z^s+z^{s-1}+\cdots+z^{-s}.
\end{equation}
$I^{(0)}$ encodes the spectrum of gauge invariant BPS operators
constructed by bi-fundamental fields.
We can easily confirm that this satisfies
(\ref{v52sym1}) and (\ref{v52sym2}).

There are six contributions to
the positive part of the index $I^{(+)}$ up to the order of $x^2$,
\begin{eqnarray}
I^{(+)}_{(\Y(1),\Y(1))}
&=&
x^{2/3}\chi_\frac{1}{2}(z_1)z_2^{1/2}
+x^{4/3}\chi_\frac{1}{2}(z_1)z_2^{1/2}
+x^{2}\chi_\frac{3}{2}(z_1)z_2^{1/2}
+\cdots,
\\
I^{(+)}_{(\Y(2),\Y(2))}
&=&
x^{4/3}\chi_1(z_1)z_2+
x^{2}(\chi_1(z_1)-1)z_2+
\cdots,
\\
I^{(+)}_{(\Y(1,1),\Y(1,1))}
&=&
x^{4/3}\chi_1(z_1)z_2+
x^{2}(\chi_1(z_1)+1)z_2+
\cdots,
\\
I^{(+)}_{(\Y(3),\Y(3))}
&=&
x^{2}(\chi_\frac{3}{2}(z_1))z_2^{3/2}+
\cdots,
\\
I^{(+)}_{(\Y(2,1),\Y(2,1))}
&=&
x^{2}(\chi_\frac{1}{2}(z_1)+\chi_\frac{3}{2}(z_1))z_2^{3/2}+\cdots,
\\
I^{(+)}_{(\Y(1,1,1),\Y(1,1,1))}
&=&
x^{2}(\chi_\frac{3}{2}(z_1))z_2^{3/2}+\cdots.
\end{eqnarray}
Monopoles with mixed charges like
$(\Y(2),\Y(1,1))$ also contribute to the index,
but they give only higher order terms.
For example,
\begin{equation}
I^{(+)}_{(\Y(2),\Y(1,1))}
=x^{14/3}z_2+\cdots.
\end{equation}
By summing up these and the trivial contribution
$I^{(+)}_{(\cdot,\cdot)}=1$, we obtain
\begin{eqnarray}
I^{(+)}&=&
1+
x^{2/3}\chi_\frac{1}{2}(z_1)z_2^{1/2}+
x^{4/3}\left(\chi_\frac{1}{2}(z_1)z_2^{1/2}+2\chi_1(z_1)z_2\right)
\nonumber\\
&&{}+x^{2}\left(\chi_\frac{3}{2}(z_1)z_2^{1/2}+2\chi_1(z_1)z_2
+\left(3\chi_\frac{3}{2}(z_1)+\chi_\frac{1}{2}(z_1)\right)z_2^{3/2}\right)+
\cdots.
\label{v52ip}
\end{eqnarray}
$I^{(-)}$ is obtained from this by the relation
(\ref{v52symc}) as
\begin{eqnarray}
I^{(-)}&=&
1+
x^{2/3}\chi_\frac{1}{2}(z_1)z_2^{-1/2}+
x^{4/3}\left(\chi_\frac{1}{2}(z_1)z_2^{-1/2}+2\chi_1(z_1)z_2^{-1}\right)\nonumber\\
&&{}+x^{2}\left(\chi_\frac{1}{2}(z_1)z_2^{-1/2}+2\chi_1(z_1)z_2^{-1}
+\left(3\chi_\frac{3}{2}(z_1)+\chi_\frac{1}{2}(z_1)\right)z_2^{-3/2}\right)+
\cdots.
\label{v52im}
\end{eqnarray}
Taking the product of 
(\ref{v52i0}), 
(\ref{v52ip}), and
(\ref{v52im}), we obtain the index,
\begin{eqnarray}
I&=&1+x^{2/3}\left(\chi_\frac{1}{2}(z_1)\chi_\frac{1}{2}(z_2)+2\right)
+x^{4/3}\left(2\chi_1(z_1)\chi_1(z_2)
+3\chi_\frac{1}{2}(z_1)\chi_\frac{1}{2}(z_2)+5\right)\nonumber\\
&&{}+x^{2}\bigg(3\chi_\frac{3}{2}(z_1)\chi_\frac{3}{2}(z_2)+
\chi_\frac{3}{2}(z_1)\chi_\frac{1}{2}(z_2)+\chi_\frac{1}{2}(z_1)\chi_\frac{3}{2}(z_2)\nonumber\\
&&\qquad\qquad{}+6\chi_1(z_1)\chi_1(z_2)
+8\chi_\frac{1}{2}(z_1)\chi_\frac{1}{2}(z_2)+10\bigg)
+\cdots.
\label{v52index}
\end{eqnarray}
This is invariant under the exchange of $z_1$ and $z_2$,
and thus invariant under the $SO(5)$ Weyl group.
We can expand
(\ref{v52index}) by $SO(5)$ characters,
\begin{eqnarray}
I&=&1+x^{2/3}\left(\chi^{SO(5)}_{(0,1)}(z_1, z_2)+1\right)
+x^{4/3}\left(2\chi^{SO(5)}_{(0,2)}(z_1,z_2)
+\chi^{SO(5)}_{(0,1)}(z_1,z_2)+2\right)\nonumber\\
&&{}+x^{2}\biggl(3\chi^{SO(5)}_{(0,3)}(z_1,z_2)
+\chi^{SO(5)}_{(2,1)}(z_1,z_2)+2\chi^{SO(5)}_{(0,2)}(z_1,z_2)\nonumber\\
&&\qquad\qquad-\chi^{SO(5)}_{(2,0)}(z_1,z_2)+3\chi^{SO(5)}_{(0,1)}(z_1,z_2)+2\biggr)
+\cdots,
\label{v52dec}
\end{eqnarray}
where $\chi_{(a_1,a_2)}^{SO(5)}$ is the character of
the representation with Dynkin label $(a_1,a_2)$.
Our convention is such that
$(1,0)$ and $(0,1)$ represent
the spinor and the vector representations
and the corresponding characters are
\begin{equation}
\chi^{SO(5)}_{(1,0)}(z_1,z_2)=
\chi_\frac{1}{2}(z_1)
+\chi_\frac{1}{2}(z_2),\quad
\chi^{SO(5)}_{(0,1)}(z_1,z_2)=
\chi_\frac{1}{2}(z_1)\chi_\frac{1}{2}(z_2)+1.
\end{equation}

When we study the relation of this index to the Kaluza-Klein spectrum
in the dual geometry,
we should compare this index to the multi-particle index.
The multi-particle index $I$ and the corresponding single particle
index $I^{\rm sp}$ on the gravity side
are related by (\ref{mpsp}).
The single particle index for
(\ref{v52dec}) is
\begin{eqnarray}
I^{\rm sp}(x,z_i)
&=&x^{2/3}\left(\chi_\frac{1}{2}(z_1)\chi_\frac{1}{2}(z_2)+2\right)
+x^{4/3}\left(\chi_1(z_1)\chi_1(z_2)
+\chi_\frac{1}{2}(z_1)\chi_\frac{1}{2}(z_2)+1\right)
\nonumber\\&&
+x^2\left(
\chi_\frac{3}{2}(z_1)\chi_\frac{3}{2}(z_2)
+\chi_1(z_1)\chi_1(z_2)
-\chi_1(z_1)
-\chi_2(z_2)+1\right)
\nonumber\\
&=&x^{2/3}(\chi_{(0,1)}^{SO(5)}(z_1,z_2)+1)
+x^{4/3}\chi_{(0,2)}^{SO(5)}(z_1,z_2)
\nonumber\\&&\quad
+x^2(\chi_{(0,3)}^{SO(5)}(z_1,z_2)-\chi_{(2,0)}^{SO(5)}(z_1,z_2))+\cdots.
\label{ispv52k1}
\end{eqnarray}

Let us next consider $k=2$ case.
$I^{(0)}$ does not depend on the Chern-Simons levels,
and is given by (\ref{v52i0}) again.
Only non-vanishing contribution to
$I^{(+)}$ up to $x^2$ is
\begin{eqnarray}
I_{(\Y(1),\Y(1))}
&=&x^{4/3}\chi_1(z_1)z_2
+x^2(\chi_1(z_1)-1)z_2+\cdots.
\end{eqnarray}
We obtain $I$ by the same way as the $k=1$ case,
\begin{equation}
I=1
+2x^{2/3}
+x^{4/3}(\chi_1(z_1)\chi_1(z_2)+4)
+x^2(3\chi_1(z_1)\chi_1(z_2)-\chi_1(z_1)-\chi_1(z_2)+7)
+\cdots.
\end{equation}
The corresponding single particle index is
\begin{equation}
I^{\rm sp}=2x^{2/3}
+x^{4/3}(\chi_1(z_1)\chi_1(z_2)+1)
+x^2(\chi_1(z_1)-1)(\chi_1(z_2)-1)
+\cdots.
\label{ispv52k2}
\end{equation}
This is consistent with the result (\ref{ispv52k1}) for $k=1$.
The dual geometry for $k=2$ is $V^{5,2}/{\bf Z}_2$.
Corresponding to the ${\bf Z}_2$ orbifolding,
(\ref{ispv52k2}) is obtained from
(\ref{ispv52k1}) by projecting away non-invariant terms under $z_2\rightarrow
e^{2\pi i}z_2$.
Still the index is invariant under the exchange of
$z_1$ and $z_2$
because of the $O(4)$ symmetry of the orbifold
$V^{5,2}/{\bf Z}_2$.

In $k=3$ case, $I^{(0)}$ is the same as above,
and non-vanishing contribution to $I^{(+)}$ up to
$x^2$ is
\begin{equation}
I_{(\Y(1),\Y(1))}(x,z_i)=x^2\chi_\frac{3}{2}(z_1)z_2^{3/2}+\cdots.
\end{equation}
The multi-particle and single-particle index are
\begin{eqnarray}
I&=&1+2x^{2/3}
+x^{4/3}(\chi_1(z_1)+4)
+x^2(
\chi_\frac{3}{2}(z_1)(z_2^{3/2}+z_2^{-3/2})
+2\chi_1(z_1)+6
)+\cdots,
\\
I^{\rm sp}&=&2x^{2/3}+x^{4/2}(\chi_1(z_1)+1)
+x^2\chi_\frac{3}{2}(z_1)(z_2^{3/2}+z_2^{-3/2})
+\cdots.
\end{eqnarray}
Again, the single particle index is obtained from
(\ref{ispv52k1})
by taking invariant terms under
$z_2\rightarrow e^{4\pi i/3}z_2$.
Now we have no symmetry between $z_1$ and $z_2$.
This is consistent with the fact that
the isometry group of $V^{5,2}/{\bf Z}_k$ with $k\geq3$
is the same as the manifest global symmetry
(\ref{v52cssym}) of the
Chern-Simons theory.

\section{Factorization in chiral theories}\label{chiral.sec}
In Section \ref{pref.sec} we proved the factorization of
the prefactor on the assumption that the theory is
vector-like and $b_0(a)$ given in (\ref{b0}) identically vanishes.
In general this is not the case and
we should treat $b_0(a)$ carefully.
In a similar way to what we used for $\epsilon_0$ and $q_{0i}$ in
Section \ref{pref.sec} we can
decompose $b_0(a)$ into three parts,
\begin{eqnarray}
b_0^{(\pm)}(a)&=&-\frac{1}{2}\sum_{\Phi_{AB}}
\sum^{(\pm)}_{\rho\in{\bf N}_A}\sum^{(\pm)}_{\rho'\in{\bf N}_B}
\nonumber\\&&\hspace{3em}
(|\rho(m)-\rho'(m)|
-|\rho(m)|-|\rho'(m)|)
(\rho(a)-\rho'(a)),\\
b'_0(a)&=&-\frac{1}{2}\sum_{\Phi_{AB}}
\sum_{\rho\in{\bf N}_A}\sum_{\rho'\in{\bf N}_B}(|\rho(m)|+|\rho'(m)|)
(\rho(a)-\rho'(a)).
\end{eqnarray}
$b_0^{(+)}$ and $b_0^{(-)}$ depend only on the positive and negative parts
of $a$, respectively.
However, unlike $\epsilon_0'$ and $q_{0i}'$,
we cannot decompose $b_0'$ into positive and negative part,
and what is worse is that $b_0'$ depends on the neutral part
$a^{(0)}$.
This fact makes the integral with respect to $a^{(0)}$ difficult.
In Section \ref{pref.sec} we took the advantage of the
$a^{(0)}$ independence of the prefactor to change the variables
$a^{(0)}$ to $\lambda_{A,n}$.
We cannot apply the same prescription if the prefactor
depends on $a^{(0)}$.

To avoid this difficulty,
we do not take account of contribution of non-diagonal
monopole operators.
Despite this ignorance of non-diagonal monopole
operators,
we expect that the index computed below
is the complete index.
The reason is as follows.
As we mentioned in Introduction,
non-diagonal monopole operators correspond
to M2-branes wrapped on two-cycles.
In general, two-cycles in the internal space are
non-BPS\cite{Benishti:2010jn}, and they cannot contribute to the index.
In the case of ${\cal N}=4$ Chern-Simons theories
the dual geometries contain shrinking two-cycles.
M2-branes wrapped on such shrinking cycles give BPS states,
and non-vanishing contribution of
non-diagonal monopole operators to the index is found\cite{Imamura:2009hc}.
However, in the examples we consider in this paper,
there are no such shrinking cycles, and thus all non-diagonal
monopole operators are expected to decouple.

For diagonal monopole operators, we can prove the factorization
of the prefactor at least for theories described by brane tilings\cite{Hanany:2005ve,Franco:2005rj,Franco:2005sm}.
The chiral theories we discuss in the following are all described by
brane tilings.
For review of brane tilings, see \cite{Kennaway:2007tq,Yamazaki:2008bt}.
See also \cite{Martelli:2008si,Hanany:2008cd,Ueda:2008hx,Imamura:2008qs}
for application of brane tilings to quiver Chern-Simons theories.

For a diagonal monopole operator with charge $m_{\rm diag}$,
we can rewrite $b'_0(a)$ as
\begin{equation}
b'_0(a)
=-\frac{1}{2}\sum_A\left(\sum_B n_{AB}\right)
\sum_{\rho\in A}(N|\rho(m)|+|m_{\rm diag}|)\rho(a).
\end{equation}
where $n_{AB}$ is the difference of the number of
chiral multiplets in $({\bf N}_A,\ol{\bf N}_B)$
and that for $(\ol{\bf N}_A,{\bf N}_B)$,
\begin{equation}
n_{AB}=\#({\bf N}_A,\ol{\bf N}_B)-\#(\ol{\bf N}_A,{\bf N}_B).
\end{equation}
In a theory described by a brane tiling,
the numbers of in-coming and out-going arrows
for each vertex are the same.
(In the context of four-dimensional quiver gauge theories,
this is necessary for the $\tr SU(N)_A^3$ gauge anomaly cancellation.)
This means
\begin{equation}
\sum_Bn_{AB}=0\quad
\forall A,
\end{equation}
and thus $b'_0(a)$ vanishes.

For the sector of diagonal monopole operators
of a theory described by a brane tiling,
simplification occurs for $\epsilon_0$ and $q_{0i}$, too.
In the sector, the relations
\begin{equation}
\sum_{A=1}^{n_G} \beta_A=0,\quad
\sum_{A=1}^{n_G}\eta_{A,i}=0.
\label{cond}
\end{equation}
hold as we will show shortly, and $\epsilon_0'$ in (\ref{eps0p})
and $q_{0i}'$ in (\ref{qp0i}) vanish.

We first prove the first equation in (\ref{cond}).
The sum of all $\beta_A$ is
\begin{equation}
\sum_{A=1}^{n_G} \beta_A=n_\Phi-n_G-\sum_\Phi\Delta_\Phi,
\label{sumaba}
\end{equation}
where $n_\Phi$ is the number of bi-fundamental chiral multiplets.
In a theory described by a brane tiling,
each field $\Phi$ appears in the
superpotential exactly twice.
Therefore, to compute $\sum_\Phi\Delta_\Phi$, we sum the Weyl weight of all terms
in the superpotential and divide it by two.
Because each term in the superpotential
has the weight $2$,
this is the number of terms in the superpotential, $n_W$,
and (\ref{sumaba}) becomes $n_\Phi-n_G-n_W$.
In a brane tiling, $n_G$, $n_\Phi$, and $n_W$ are
the numbers of faces, edges, and vertices, respectively,
and (\ref{sumaba}) is nothing but
the opposite of Euler's characteristic
of the surface on which the tiling is drawn.
Because the brane tiling is always drawn on a torus,
it always vanishes.

Next, let us consider
the second equation in (\ref{cond}).
With the definition of $\eta_{A,i}$ in (\ref{fa}),
we obtain
\begin{equation}
\sum_{A=1}^{n_G}\eta_{A,i}=\sum_\Phi F_i(\Phi)
=\frac{1}{2}\sum_{I=1}^{n_W}\sum_{\Phi\in I}F_i(\Phi).
\label{allfa}
\end{equation}
The index $I$ labels terms in the superpotential,
and $\Phi\in I$ means fields contained in the $I$-th term
in the superpotential.
In the second equality we again used the fact that every chiral multiplet
appears in the superpotential exactly twice.
By definition, flavor charges of the superpotential vanish, and
the inner summation on the right hand side
in (\ref{allfa}) vanishes.
Therefore the second equation in (\ref{cond}) holds.

Combining results in this section,
the positive part of the prefactor for diagonal monopole operators
is given by
\begin{equation}
P^{(+)}=e^{-S_{\rm CS}^{(+)}}
e^{ib_0^{(+)}(a)}
x^{\epsilon_0^{(+)}}
z_i^{q_{0i}^{(+)}}.
\end{equation}

\section{Example 2: $Q^{1,1,1}$}\label{q111.sec}
The coset space
\begin{equation}
\frac{SU(2)\times SU(2)\times SU(2)}{U(1)\times U(1)}
\end{equation}
is called $Q^{1,1,1}$, and has
the isometry
\begin{equation}
SU(2)_1\times SU(2)_2\times SU(2)_3\times U(1).
\end{equation}
The last $U(1)$ factor is identified with the R-symmetry.
Quiver Chern-Simons theories
corresponding to this manifold are proposed in \cite{Franco:2008um}.
See also \cite{Benishti:2010jn,Franco:2009sp} for further investigation.

\begin{figure}[htb]
\centerline{\includegraphics{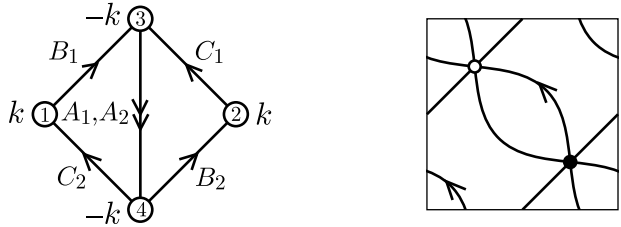}}
\caption{The quiver diagram and the brane tiling
for a theory dual to $Q^{1,1,1}/{\bf Z}_k$.}
\label{q111quiv}
\end{figure}

This manifold is toric, and dual theories are
described by brane tilings.
We first consider the theory shown in Fig.\ \ref{q111quiv}.
The arrows on the edges in the brane tiling represent
gradient of the edges in the corresponding
brane crystal\cite{Lee:2006hw,Lee:2007kv,Kim:2007ic}
obtained in the way proposed in \cite{Imamura:2008qs}.
The superpotential of the Chern-Simons theory is
\begin{equation}
W=\tr(\epsilon^{ij}C_2B_1A_iB_2C_1A_j).
\end{equation}
When the Chern-Simons levels are $(k,k,-k,-k)$,
the corresponding geometry is $Q^{1,1,1}/{\bf Z}_k$
where ${\bf Z}_k$ is a subgroup of the diagonal $SU(2)$ in
$SU(2)_2\times SU(2)_3$,
and break it down to $U(1)_F\times U(1)_B$.
The manifest global symmetry of this theory is
\begin{equation}
SU(2)\times U(1)_F\times U(1)_B\times U(1)_R.
\end{equation}
$SU(2)$ is the symmetry acting on $A_i$.
Note that the action does not have $SU(2)$ symmetries
rotating $B_i$ and $C_i$.
We define three generators $F_i$ ($i=1,2,3$)
of the Cartan part of the flavor symmetry $SU(2)\times U(1)_F\times U(1)_B$.
$F_2+F_3$ and $F_2-F_3$ generate $U(1)_B$ and $U(1)_F$,
respectively.
The charge assignments are shown in 
Table \ref{table:q111}.
\begin{table}[htb]
\caption{Charge assignments of R and flavor symmetries
of the theory in Fig.\ \ref{q111quiv} are shown.}
\label{table:q111}
\begin{center}
\begin{tabular}{ccccccc}
\hline
\hline
      & $A_1$ & $A_2$ & $B_1$ & $B_2$ & $C_1$ & $C_2$ \\
\hline
$\Delta$ & $1-2h$ & $1-2h$ & $h$ & $h$ & $h$ & $h$ \\
$F_1$ & $\frac{1}{2}$ & $-\frac{1}{2}$ & $0$ & $0$ & $0$ & $0$ \\
$F_2$ & $0$ & $0$ & $\frac{1}{2}$ & $-\frac{1}{2}$ & $0$ & $0$ \\
$F_3$ & $0$ & $0$ & $0$ & $0$ & $\frac{1}{2}$ & $-\frac{1}{2}$ \\
\hline
\end{tabular}
\end{center}
\end{table}
From only the symmetry of the action and the marginality of
the superpotential the R-charges of the chiral fields are not fixed.
We have an ambiguity of an unknown parameter $h$.
The index does not depend on $h$, and we do not need to
determine it.

The $SU(2)$ symmetry of the action
guarantees that
the index satisfy
\begin{equation}
I(x,z_1,z_2,z_3)=I(x,z_1^{-1},z_3,z_2),
\label{q111isym1}
\end{equation}
and the index is
expanded by $SU(2)$ characters $\chi_s(z_1)$.
This theory also has ${\bf Z}_2$ symmetry which exchanges
$B_i$ and $C_i$,
and the index satisfies the relation
\begin{equation}
I(x,z_1,z_2,z_3)=I(x,z_1,z_3,z_2).
\label{q111isym2}
\end{equation}
Another relation follows from the charge conjugation symmetry.
If we apply the charge conjugation to the theory,
all arrows are reversed and
the resulting diagram is obtained
by rotating the original one by 180 degrees.
This exchanges $B_1$ and $B_2$,
and $C_1$ and $C_2$.
This implies the relation
\begin{equation}
I^{(-)}(x,z_1,z_2,z_3)=I^{(+)}(x,z_1,z_2^{-1},z_3^{-1}),
\end{equation}
and the symmetry of the index,
\begin{equation}
I(x,z_1,z_2,z_3)
=I(x,z_1,z_2^{-1},z_3^{-1}).
\label{q111isym3}
\end{equation}

If the flavor symmetry is enhanced to $SU(2)^3$ in $k=1$ case,
$I(x,z_i)$ must have a larger symmetry.
It should be invariant under Weyl Group of three $SU(2)$,
which inverses three $z_i$ independently.
We also expect the permutation symmetry among $z_i$.
Let us confirm this symmetry enhancement by computing the index
in $k=1$ case.

The neutral part of the index is
\begin{equation}
I^{(0)}(x,z_i)=1+
\chi_\frac{1}{2}(z_1)
\left(\frac{z_2^{1/2}}{z_3^{1/2}}+\frac{z_3^{1/2}}{z_2^{1/2}}\right)
x+
\left[
\chi_1(z_1)\left(2\frac{z_2}{z_3}+1+2\frac{z_3}{z_2}\right)
-3
\right]
x^2+\cdots.
\label{q111i0}
\end{equation}

There are three contributions to $I^{(+)}$ up to the order of $x^2$,
\begin{eqnarray}
I^{(+)}_{(\Y(1),\Y(1),\Y(1),\Y(1))}(x,z_i)
&=&\chi_\frac{1}{2}(z_1)z_2^{-1/2}z_3^{-1/2}x
+(\chi_1(z_1)-1)(z_2^{-1}+z_3^{-1})x^2
+\cdots,
\nonumber\\
I^{(+)}_{(\Y(2),\Y(2),\Y(2),\Y(2))}(x,z_i)
&=&\chi_1(z_1)z_2^{-1}z_3^{-1}x^2
+\cdots,
\nonumber\\
I^{(+)}_{(\Y(1,1),\Y(1,1),\Y(1,1),\Y(1,1))}(x,z_i)
&=&\chi_1(z_1)z_2^{-1}z_3^{-1}x^2+\cdots.
\end{eqnarray}
Summing up these three and the trivial one
$I^{(+)}_{(\cdot,\cdot,\cdot,\cdot)}=1$,
we obtain
\begin{equation}
I^{(+)}=1+\chi_\frac{1}{2}(z_1)z_2^{-1/2}z_3^{-1/2}x
+\left[(\chi_1(z_1)-1)(z_2^{-1}+z_3^{-1})
+2\chi_1(z_1)z_2^{-1}z_3^{-1}\right]x^2
+\cdots.
\label{q111ip}
\end{equation}
This satisfies
(\ref{q111isym1}) and
(\ref{q111isym2}).
We obtain $I^{(-)}$ by the relation
(\ref{q111isym3}),
\begin{equation}
I^{(-)}=1+\chi_\frac{1}{2}(z_1)z_2^{1/2}z_3^{1/2}x
+\left[(\chi_1(z_1)-1)(z_2+z_3)
+2\chi_1(z_1)z_2z_3\right]x^2
+\cdots.
\label{q111im}
\end{equation}
As the product of
(\ref{q111i0}), (\ref{q111ip}), and (\ref{q111im})
we obtain the index
\begin{equation}
I(x,z_i)
=1
+\chi_{\frac{1}{2}}(z_1)\chi_{\frac{1}{2}}(z_2)\chi_{\frac{1}{2}}(z_3)x
+(2\chi_1(z_1)\chi_1(z_2)\chi_1(z_3)-2)x^2
+\cdots.
\label{indexq111}
\end{equation}
This is invariant under the Weyl reflections
$z_2\rightarrow z_2^{-1}$
and $z_3\rightarrow z_3^{-1}$,
and under permutations among $z_i$.
This is precisely what we expect from the isometry of $Q^{1,1,1}$.
The single particle index for (\ref{indexq111}) is
\begin{eqnarray}
&&
I^{\rm sp}(x,z_i)=
\chi_\frac{1}{2}(z_1)
\chi_\frac{1}{2}(z_2)
\chi_\frac{1}{2}(z_3)x
\nonumber\\&&\hspace{2em}
+
\left(
\chi_1(z_1)
\chi_1(z_2)
\chi_1(z_3)
-\chi_1(z_1)
-\chi_1(z_2)
-\chi_1(z_3)
-2
\right)x^2+\cdots.
\label{q111sp}
\end{eqnarray}

We also consider $k=2$ case.
Only non-vanishing contribution to $I^{(+)}$ up to the order of $x^2$ is
\begin{equation}
I_{(\Y(1),\Y(1),\Y(1),\Y(1))}=x^2\frac{\chi_1(z_1)}{z_2z_3}+\cdots,
\end{equation}
and the index $I$ and the corresponding single particle index $I^{\rm sp}$ are
\begin{eqnarray}
I&=&1+x\chi_1(z_1)\left(\frac{z_2^{1/2}}{z_3^{1/2}}+\frac{z_3^{1/2}}{z_2^{1/2}}\right)
\nonumber\\&&
+x^2\left[\chi_1(z_1)\left(z_2z_3+\frac{1}{z_2z_3}
+2\frac{z_2}{z_3}
+2\frac{z_3}{z_2}+1
\right)-3\right]\cdots,\\
I^{\rm sp}
&=&x\chi_1(z_1)\left(\frac{z_2^{1/2}}{z_3^{1/2}}+\frac{z_3^{1/2}}{z_2^{1/2}}\right)
\nonumber\\&&
+x^2\left[\chi_1(z_1)\left(
z_2z_3+\frac{1}{z_2z_3}
+\frac{z_2}{z_3}+\frac{z_3}{z_2}
\right)-4\right]+\cdots.
\label{eq106}
\end{eqnarray}
Because the dual geometry is the orbifold $Q^{1,1,1}/{\bf Z}_k$,
(\ref{eq106}) should be obtained from (\ref{q111sp}) by the projection
associated with $(z_1,z_2,z_3)\rightarrow(z_1,e^{\pi i}z_2,e^{\pi i}z_3)$.
We can easily confirm that this is actually the case.

There is another realization of $Q^{1,1,1}$
by a quiver Chern-Simons theory\cite{Franco:2008um}.
It is described by the same quiver diagram as the previous one
and has the same superpotential,
but has different Chern-Simons levels $( k,-k,0,0 )$.
\begin{figure}[htb]
\centerline{\includegraphics{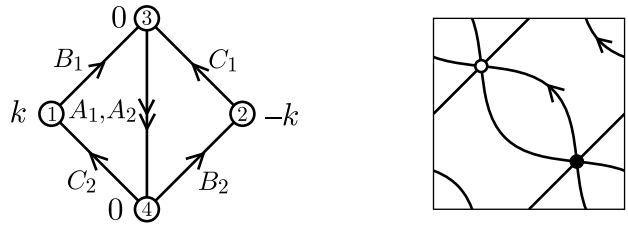}}
\caption{The quiver diagram and the brane tiling
for a theory dual to $Q^{1,1,1}/{\bf Z}_k$.
The Chern-Simons levels are
different from Figure \ref{q111quiv2.eps}.}
\label{q111quiv2.eps}
\end{figure}

The charge assignments are shown in 
Table \ref{table:q111_2}. 
\begin{table}[htb]
\caption{Charge assignments of R and flavor symmetries
of the theory in Fig.\ \ref{q111quiv2.eps} are shown.}
\label{table:q111_2}
\begin{center}
\begin{tabular}{ccccccc}
\hline
\hline
      & $A_1$ & $A_2$ & $B_1$ & $B_2$ & $C_1$ & $C_2$ \\
\hline
$\Delta$ & $1-2h$ & $1-2h$ & $h$ & $h$ & $h$ & $h$ \\
$F_1$ & $\frac{1}{2}$ & $-\frac{1}{2}$ & $0$ & $0$ & $0$ & $0$ \\
$F_2$ & $0$ & $0$ & $\frac{1}{2}$ & $0$ & $-\frac{1}{2}$ & $0$ \\
$F_3$ & $0$ & $0$ & $0$ & $\frac{1}{2}$ & $0$ & $-\frac{1}{2}$ \\
\hline
\end{tabular}
\end{center}
\end{table}
Due to the different Chern-Simons levels,
the $U(1)_B$ charge of this theory rotates fields in
a different way from the previous one.
We define $F_2$ and $F_3$ so that $F_2+F_3$ again generates
the baryonic symmetry.
The manifest flavor symmetry of the action
guarantees
\begin{equation}
I(x,z_1,z_2,z_3)
=I(x,z_1^{-1},z_2,z_3),
\end{equation}
and the charge conjugation gives
the relation
\begin{equation}
I(x,z_1,z_2,z_3)
=I(x,z_1,z_3^{-1},z_2^{-1}).
\end{equation}

We computed the index for $k=1,2$, and we obtain the same results
for $I^{(0)},I^{(+)}$ and $I^{(-)}$ as the previous theory for $Q^{1,1,1}$.

\section{Example 3: $Q^{2,2,2}$}\label{q222.sec}
The homogeneous Sasaki-Einstein space $Q^{2,2,2}$
is defined as the orbifold $Q^{1,1,1}/{\bf Z}_2$.
The generator of the orbifold group ${\bf Z}_2$ is $e^{\pi i(R+2j_3)}$.
The isometry group of $Q^{2,2,2}$ is the same as $Q^{1,1,1}$.

\begin{figure}[htb]
\centerline{\includegraphics{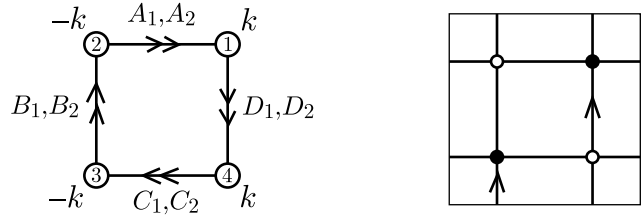}}
\caption{The quiver diagram and the brane tiling for a theory dual to $Q^{2,2,2}/{\bf Z}_k$.}
\label{q222quiv}
\end{figure}

Dual Chern-Simons theories for $Q^{2,2,2}$ are proposed in \cite{Franco:2008um}.
We first consider the theory shown in 
Fig.\ \ref{q222quiv}.
This theory has the superpotential
\begin{equation}
W=\epsilon^{ik}\epsilon^{jl}\tr(A_iB_jC_kD_l).
\end{equation}
The manifest global symmetry of this theory is
\begin{equation}
SU(2)\times SU(2)\times U(1)_B\times U(1)_R.
\end{equation}
The first $SU(2)$ acts on $A_i$ and $C_i$,
and the second on $B_i$ and $D_i$.
Let $F_1$ and $F_2$ be the generators of two $SU(2)$ groups,
and $F_3$ be the generator of the baryonic $U(1)$.
The conformal dimensions of fields and the charge assignments are shown
in Table \ref{table:q222}.
\begin{table}[htb]
\caption{The R-charge assignment and charge assignments
of the theory in Fig.\ \ref{q222quiv} are shown.}
\label{table:q222}
\begin{center}
\begin{tabular}{ccccccccc}
\hline
\hline
& $A_1$ & $A_2$ & $B_1$ & $B_2$ & $C_1$ & $C_2$ & $D_1$ & $D_2$ \\
\hline
$\Delta$ & $h$ & $h$ & $1-h$ & $1-h$ &
           $h$ & $h$ & $1-h$ & $1-h$ \\
$F_1$ & $\frac{1}{2}$ & $-\frac{1}{2}$ & $0$ & $0$ & $\frac{1}{2}$ & $-\frac{1}{2}$ & $0$ & $0$ \\
$F_2$ & $0$ & $0$ & $\frac{1}{2}$ & $-\frac{1}{2}$ & $0$ & $0$ & $\frac{1}{2}$ & $-\frac{1}{2}$ \\
$F_3$ & $\frac{1}{2}$ & $\frac{1}{2}$ & $0$ & $0$ & $-\frac{1}{2}$ & $-\frac{1}{2}$ & $0$ & $0$ \\
\hline
\end{tabular}
\end{center}
\end{table}
The Weyl group of the
flavor symmetry guarantees the relations
\begin{equation}
I(x,z_1,z_2,z_3)
=I(x,z_1^{-1},z_2,z_3)
=I(x,z_1,z_2^{-1},z_3).
\end{equation}
We also have
\begin{equation}
I(x,z_1,z_2,z_3)=I(x,z_1,z_2,z_3^{-1})
\label{q222crel}
\end{equation}
from the charge conjugation symmetry.
These generate the Weyl group of
$SU(2)^3$, but it is non-trivial
whether the index respects the permutation symmetry among $z_i$ when $k=1$.

Let us compute the index for $k=1$.
The neutral part is
\begin{equation}
I^{(0)}=1+x^2
\left[(\chi_1(z_1)-1)(\chi_1(z_2)-1)-4\right]+\cdots,
\label{q222i0}
\end{equation}
and the only monopole contributions to $I^{(+)}$ up to $x^2$ is
\begin{equation}
I^{(+)}_{(\Y(1),\Y(1),\Y(1),\Y(1))}
=x^2\left[\chi_1(z_1)\chi_1(z_2)-1\right]z_3+\cdots.
\end{equation}
The whole index $I$ and the corresponding single particle index $I^{\rm sp}$
are
\begin{eqnarray}
I&=&1+\left[\chi_1(z_1)\chi_1(z_2)\chi_1(z_3)
-\chi_1(z_1)-\chi_1(z_2)-\chi_1(z_3)-2\right]x^2+\cdots,
\label{iq222}\\
I^{\rm sp}&=&
\left(\chi_1(z_1)\chi_1(z_2)\chi_1(z_3)
-\chi_1(z_1)-\chi_1(z_2)-\chi_1(z_3)-2
\right)x^2+\cdots.
\label{ispq222}
\end{eqnarray}
These have expected permutation symmetry.
Because $Q^{2,2,2}$ is the ${\bf Z}_2$ orbifold
of $Q^{1,1,1}$,
(\ref{ispq222}) should be related to (\ref{q111sp})
by the orbifold projection.
The orbifold group ${\bf Z}_2$ is generated by
$e^{\pi i(R+2j_3)}$.
On BPS operators saturating the BPS bound (\ref{BPSbound})
this parity is equal to $e^{\pi i(\Delta+j_3)}$, and
flips the sign of $x$.
(\ref{ispq222})
is what is obtained from (\ref{q111sp}) by the projection
associated with this ${\bf Z}_2$ flip $x\rightarrow -x$.

\begin{figure}[htb]
\centerline{\includegraphics{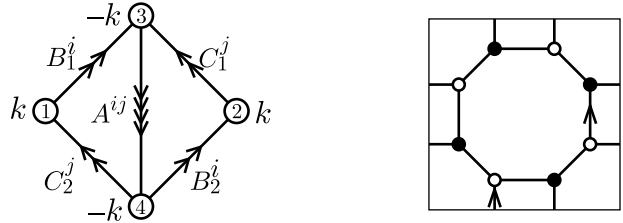}}
\caption{The quiver diagram and the brane tiling for a
dual theory to $Q^{2,2,2}/{\bf Z}_k$ are shown.}
\label{q222quiv2.eps}
\end{figure}

There is another theory proposed as a dual to $Q^{2,2,2}$\cite{Franco:2008um}.
(Fig.\ \ref{q222quiv2.eps})
The superpotential of this theory is
\begin{equation}
W=\tr(
\epsilon_{ij}\epsilon_{kl}A^{ik}B_1^jC_2^l
-\epsilon_{ij}\epsilon_{kl}A^{ik}B_2^jC_1^l),
\end{equation}
and the manifest global symmetry of the action is
\begin{equation}
SU(2)_1\times SU(2)_2\times U(1)_B\times U(1)_R.
\end{equation}
We define $F_1$, $F_2$, and $F_3$ as generators of
the $SU(2)_1$ Cartan,
the $SU(2)_2$ Cartan,
and $U(1)_B$, respectively.
\begin{table}[htb]
\caption{The R and flavor charge assignments of the theory in Fig.\ \ref{q222quiv2.eps} are shown.}
\label{table:q2222}
\begin{center}
\begin{tabular}{cccccc}
\hline
\hline
      & $A^{ij}$ & $B_1^i$ & $B_2^i$ & $C_1^j$ & $C_2^j$ \\
\hline
$\Delta$ & $2-2h$ & $h$ & $h$ \\
$F_1$ & $\pm\frac{1}{2}$ & $\pm\frac{1}{2}$ & $\pm\frac{1}{2}$ & $0$ & $0$ \\
$F_2$ & $\pm\frac{1}{2}$ & $0$ & $0$ & $\pm\frac{1}{2}$ & $\pm\frac{1}{2}$ \\
$F_3$ & $0$ & $\frac{1}{2}$ & $-\frac{1}{2}$ & $\frac{1}{2}$ & $-\frac{1}{2}$ \\
\hline
\end{tabular}
\end{center}
\end{table}
The manifest $SU(2)_1\times SU(2)_2$ symmetry guarantees
the relations
\begin{equation}
I(x,z_1,z_2,z_3)
=I(x,z_1^{-1},z_2,z_3)
=I(x,z_1,z_2^{-1},z_3),
\end{equation}
and the charge conjugation guarantees
\begin{equation}
I(x,z_1,z_2,z_3)
=I(x,z_1,z_2,z_3^{-1}).
\end{equation}
The neutral part is the same as (\ref{q222i0}),
and the only non-vanishing contribution to $I^{(+)}$ up to
the order of $x^2$ is
\begin{equation}
I_{(\Y(1),\Y(1),\Y(1),\Y(1))}=x^2[\chi_1(z_1)\chi_1(z_2)-1]z_3^{-1}.
\end{equation}
The indices $I$ and $I^{\rm sp}$ are identical to
(\ref{iq222}) and (\ref{ispq222}), respectively.

\section{Example 4: $M^{1,1,1}$}\label{m111.sec}
$M^{1,1,1}$, which is also often called $M^{3,2}$, is the coset
\begin{equation}
\frac{SU(3)\times SU(2)\times U(1)}{SU(2)\times U(1)\times U(1)},
\end{equation}
and has the isometry
\begin{equation}
SU(3)\times SU(2)\times U(1).
\end{equation}
The $U(1)$ factor is identified with the $U(1)_R$ symmetry.

\begin{figure}[htb]
\centerline{\includegraphics{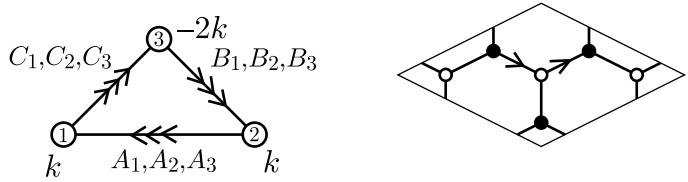}}
\caption{The quiver diagram for a theory dual to $M^{1,1,1}/{\bf Z}_k$.}
\label{m111quiv}
\end{figure}

A quiver Chern-Simons theory dual to $M^{1,1,1}$ is proposed in
\cite{Martelli:2008si,Hanany:2008cd}.
It is $U(N)^3$ Chern-Simons theory shown in Fig.\ \ref{m111quiv}.
The superpotential is
\begin{equation}
W=\epsilon^{ijk}\tr(A_iB_jC_k).
\end{equation}
The manifest global symmetry of this theory is
\begin{equation}
SU(3)\times U(1)_B\times U(1)_R,
\end{equation}
where $SU(3)$ rotates $A_i$, $B_i$, and $C_i$ simultaneously.
$U(1)_B$ is the baryonic symmetry acting on $B_i$ and $C_i$ with opposite
charges.
We introduce $F_1$ and $F_2$ as $SU(3)$ Cartan generators,
and $F_3$ as $U(1)_B$ charge.
The charge assignments are shown
in Table \ref{table:m111}.
\begin{table}[htb]
\caption{Charge assignments of R and flavor symmetries for the $M^{1,1,1}$ theory are shown.}
\label{table:m111}
\begin{center}
\begin{tabular}{cccccccccc}
\hline
\hline
      & $A_1$ & $A_2$ & $A_3$ & $B_1$ & $B_2$ & $B_3$ & $C_1$ & $C_2$ & $C_3$ \\
\hline
$\Delta$ & $2-2h$ & $2-2h$ & $2-2h$ & $h$ & $h$ & $h$ & $h$ & $h$ & $h$ \\
$F_1$ & $1$ & $-1$ & $0$ & $1$ & $-1$ & $0$ & $1$ & $-1$ & $0$ \\
$F_2$ & $0$ & $1$ & $-1$ & $0$ & $1$ & $-1$ & $0$ & $1$ & $-1$ \\
$F_3$ & $0$ & $0$ & $0$ & $\frac{1}{2}$ & $\frac{1}{2}$ & $\frac{1}{2}$ & $-\frac{1}{2}$ & $-\frac{1}{2}$ & $-\frac{1}{2}$ \\
\hline
\end{tabular}
\end{center}
\end{table}

Due to the manifest $SU(3)$ symmetry,
the index is invariant under the $SU(3)$ Weyl group generated by the relations
\begin{equation}
I(x,z_1,z_2,z_3)
=I(x,z_2^{-1},z_1^{-1},z_3)
=I(x,z_1,z_1z_2^{-1},z_3).
\label{su3weyl}
\end{equation}
The charge conjugation
gives the relation
\begin{equation}
I^{(-)}(x,z_1,z_2,z_3)
=I^{(+)}(x,z_1,z_2,z_3^{-1}),
\end{equation}
between $I^{(+)}$ and $I^{(-)}$,
and the symmetry of the index
\begin{equation}
I(x,z_1,z_2,z_3)
=I(x,z_1,z_2,z_3^{-1}).
\label{m111cc}
\end{equation}
Disappointingly,
the Weyl group of the
full flavor symmetry $SU(3)\times SU(2)$
is exhausted by the relations generated by
(\ref{su3weyl}) and (\ref{m111cc}), and thus
computation of the index and simple symmetry analysis
does not give extra information about the symmetry
enhancement from $U(1)_B$ to $SU(2)$.
We give some results in $k=1$ case just for reference.
The neutral part of the index up to $x^2$ is
\begin{eqnarray}
I^{(0)}
&=&1
+
\left(z_1^3+\frac{z_2^3}{z_1^3}+\frac{1}{z_2^3}-3\right)
x^2
+\cdots
\nonumber\\
&=&1
+
(\chi^{SU(3)}_{(3,0)}(z_1,z_2)-\chi^{SU(3)}_{(1,1)}(z_1,z_2)-2)x^2
+\cdots,
\end{eqnarray}
where $\chi^{SU(3)}_{(a_1,a_2)}$ is the
$SU(3)$ character of
the representation with
Dynkin label $(a_1,a_2)$.
Our definition here is such that for the fundamental
and the anti-fundamental representations,
\begin{equation}
\chi^{SU(3)}_{(1,0)}(z_1,z_2)=z_1+\frac{z_2}{z_1}+\frac{1}{z_2},\quad
\chi^{SU(3)}_{(0,1)}(z_1,z_2)=\frac{1}{z_1}+\frac{z_1}{z_2}+z_2.
\label{su3ch}
\end{equation}

The only monopole contribution to $I^{(+)}$ up to order $x^2$ is
\begin{eqnarray}
I_{(\Y(1),\Y(1),\Y(1))}^{(+)}
&=&\left(z_1+\frac{z_2}{z_1}+\frac{1}{z_2}\right)
\left(z_1^2+\frac{z_2^2}{z_1^2}+\frac{1}{z_2^2}\right)z_3x^2
+\cdots
\nonumber\\
&=&(\chi_{(3,0)}^{SU(3)}(z_1,z_2)-1)z_3x^2+\cdots.
\end{eqnarray}
The index $I$ and the corresponding $I^{\rm sp}$ are
\begin{eqnarray}
I
&=&1
+
[(\chi_{(3,0)}^{SU(3)}(z_1,z_2)-1)\chi_1(z_3)
-\chi_{(1,1)}^{SU(3)}(z_1,z_2)-1]x^2
+\cdots,\\
I^{\rm sp}
&=&
[(\chi_{(3,0)}^{SU(3)}(z_1,z_2)-1)\chi_1(z_3)
-\chi_{(1,1)}^{SU(3)}(z_1,z_2)-1]x^2
+\cdots.
\end{eqnarray}

\section{Adding flavors}\label{flavor.sec}
Let us extend the formula obtained in previous sections to
the theories with flavors belonging to fundamental and anti-fundamental representations\cite{Ammon:2009wc,Gaiotto:2009tk,Benini:2009qs,Jafferis:2009th}.

The contribution of chiral multiplets in (anti-)fundamental representations
to the letter index is
obtained from (\ref{fchdef}) by setting $\rho'=0$ ($\rho=0$).
The contribution of fundamental representations is
\begin{eqnarray}
f^{\bf N}_{\rm chiral}(e^{ia},x,z_i)
&=&
\sum_{\Phi_{A0}}
\sum_{\rho\in{\bf N}_A}
\frac{x^{|\rho(m)|}}{1-x^2}
\nonumber\\&&\hspace{2em}
\left(
e^{i\rho(a)}z_i^{F_i(\Phi)}x^{\Delta(\Phi)}
-e^{-i\rho(a)}z_i^{-F_i(\Phi)}x^{2-\Delta(\Phi)}
\right)\nonumber\\
&=&
\sum_{\Phi_{A0}}
\left(
\lambda_{A,+1}\frac{z_i^{F_i(\Phi)}x^{\Delta(\Phi)}}{1-x^2}
-\lambda_{A,-1}\frac{z_i^{-F_i(\Phi)}x^{2-\Delta(\Phi)}}{1-x^2}
\right),
\label{ffund}
\end{eqnarray}
where $\Phi_{A0}$ represents chiral multiplets
belonging to the $U(N)_A$ fundamental representation ${\bf N}_A$.
The contribution of anti-fundamental representations
is
\begin{eqnarray}
&&f^{\ol{\bf N}}_{\rm chiral}(e^{ia},x,z_i)
=
\sum_{\Phi_{0B}}
\left(
\lambda_{B,-1}\frac{z_i^{F_i(\Phi)}x^{\Delta(\Phi)}}{1-x^2}
-\lambda_{B,+1}\frac{z_i^{-F_i(\Phi)}x^{2-\Delta(\Phi)}}{1-x^2}
\right).
\label{fafund}
\end{eqnarray}
$\Phi_{0B}$ represents chiral multiplets belonging to the
$U(N)_B$ anti-fundamental representation $\ol{\bf N}_B$

(\ref{ffund}) and (\ref{fafund})
depend on $a^{(0)}$ through $\lambda_{A,m}$
defined in (\ref{lambdadef}),
and we include this contribution to the definition of $f'$.
There is no change in the definition of $f^{(\pm)}$.
Only bi-fundamental fields contribute to $f^{(\pm)}$.

Let us define $V_A$ and $W_A$ by
\begin{equation}
f_{\rm chiral}^{\bf N}
+f_{\rm chiral}^{\ol{\bf N}}
=\sum_{A=1}^{n_G}(V_A\lambda_{A,-1}+\lambda_{A,1}W_A).
\label{linearlambda}
\end{equation}
The neutral part of index $I^{(0)}$ is defined by
Gaussian integral including (\ref{linearlambda}) in the
exponent,
\begin{eqnarray}
I^{(0)}&=&\int d\lambda
\exp\sum_{n=1}^\infty
\frac{1}{n}\Bigg(-\sum_{A,B}\lambda_{A,n}M_{A,B}(\cdot^n)\lambda_{B,-n}
\nonumber\\&&
\hspace{9em}+\sum_AV_A(\cdot^n)\lambda_{A,-n}
+\sum_A\lambda_{A,n}W_A(\cdot^n)
\Bigg)
\nonumber\\
&=&\prod_{n=1}^\infty
\frac{1}{\det M_{A,B}(\cdot^n)}\exp\left(
\frac{1}{n}\sum_{A,B}V_A(\cdot^n)(M^{-1}(\cdot^n))_{A,B}W_B(\cdot^n)
\right)
\label{i0fund}
\end{eqnarray}
Compared to (\ref{neutralint2}),
(\ref{i0fund}) has the extra exponential factor.
This contributes to the single particle index by
\begin{equation}
\sum_{A,B}V_A(x,z_i)(M^{-1}(x,z_i))_{A,B}W_B(x,z_i).
\end{equation}

The introduction of flavors changes
the zero-point contributions to
the energy, the flavor charges, and the gauge charges.
The extra contributions are
\begin{eqnarray}
\epsilon_0^{\rm fund}
&=&
\frac{1}{2}\sum_{\Phi_{A0},\Phi_{0A}}
\sum_{\rho\in{\bf N}_A}|\rho(m)|(1-\Delta(\Phi)),
\label{ep0fund}\\
q_{0i}^{\rm fund}
&=&-\frac{1}{2}\sum_{\Phi_{A0},\Phi_{0A}}
\sum_{\rho\in{\bf N}_A}|\rho(m)|F_i(\Phi),
\label{q0ifund}
\\
b_0^{\rm fund}(a)
&=&
-\frac{1}{2}\sum_{\Phi_{A0}}
\sum_{\rho\in{\bf N}_A}|\rho(m)|\rho(a)
+\frac{1}{2}\sum_{\Phi_{0A}}
\sum_{\rho\in{\bf N}_A}|\rho(m)|\rho(a).
\label{b0fund}
\end{eqnarray}
For vector-like theories, $b_0(a)$ identically vanishes.
These do not depend on $a^{(0)}$, and are added to
the definition of $\epsilon^{(\pm)}$,
$q_{0i}^{(\pm)}$, and $b_0^{(\pm)}(a)$.

\section{Example 5: $N^{0,1,0}$}\label{n010.sec}
As an example of a theory with flavors,
let us consider a flavored ABJM model,
which is proposed as a dual theory to $N^{0,1,0}$\cite{Gaiotto:2009tk}.

$N^{0,1,0}$ is the homogeneous manifold defined as the coset
\begin{equation}
SU(3)/U(1),
\end{equation}
and its isometry is
\begin{equation}
SU(3)\times SU(2).
\label{n010iso}
\end{equation}
This manifold is hyper-K\"ahler and respects ${\cal N}=3$ supersymmetry.
The $SU(2)$ factor
in (\ref{n010iso}) is the R-symmetry rotating three supercharges.
We use only two of them for the localization.

\begin{figure}[htb]
\centerline{\includegraphics{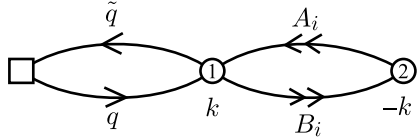}}
\caption{The quiver diagram of a flavored ABJM model dual to
$N^{0,1,0}$ is shown.
The box represents a global $U(1)$ symmetry.}
\label{n010quiv.eps}
\end{figure}

The theory dual to $N^{0,1,0}$ is defined by adding one vector-like flavor $(q,\wt q)$ to the ABJM model, which has gauge group $U(N)_1\times U(N)_2$.
$q$ and $\wt q$ belong to the fundamental and the anti-fundamental representations of $U(N)_1$.
The quiver diagram of this theory is shown in 
Fig.\ \ref{n010quiv.eps}.
The superpotential is
\begin{equation}
W=\tr[(\epsilon^{ij}A_iB_j+q\wt q)^2-(\epsilon^{ij}B_iA_j)^2]
\end{equation}
The introduction of the flavor breaks the
${\cal N}=6$ supersymmetry of the ABJM model to ${\cal N}=3$.
Due to the non-abelian R-symmetry $SU(2)_R$
the R-charges are protected from quantum corrections
and all chiral multiplets have the canonical R-charge $1/2$.
The manifest global symmetry of the theory is
\begin{equation}
SU(2)\times U(1)_B\times SU(2)_R.
\end{equation}
$SU(2)$ rotates $A_i$ and $B_i$ as doublets.
The dual geometry of this model is $N^{0,1,0}/{\bf Z}_k$.
We would like to
confirm that the flavor symmetry $SU(2)\times U(1)_B$
is enhanced to $SU(3)$ in $k=1$ case.
Let $F_1$ and $F_2$ be the generators of $SU(2)$ Cartan and the
baryonic symmetry, respectively.
The charge assignments are shown in Table \ref{table:n010}.
\begin{table}[htb]
\caption{The charge assignments of R and flavor symmetries
in the $N^{0,1,0}$ theory are shown.}
\label{table:n010}
\begin{center}
\begin{tabular}{ccccccc}
\hline
\hline
&$q$&$\wt q$&$A_1$&$A_2$&$B_1$&$B_2$\\
\hline
$\Delta$ & $\frac{1}{2}$ & $\frac{1}{2}$ & $\frac{1}{2}$ & $\frac{1}{2}$ & $\frac{1}{2}$ & $\frac{1}{2}$ \\
$F_1$&$0$&$0$&$\frac{1}{2}$&$-\frac{1}{2}$&$\frac{1}{2}$&$-\frac{1}{2}$\\
$F_2$&$1$&$-1$&$1$&$1$&$-1$&$-1$\\
\hline
\end{tabular}
\end{center}
\end{table}
The manifest $SU(2)$ symmetry
guarantees the relation
\begin{equation}
I(x,z_1,z_2)=I(x,z_1^{-1},z_2).
\end{equation}
From the charge conjugation symmetry, we obtain the relation
\begin{equation}
I^{(-)}(x,z_1,z_2)=I^{(+)}(x,z_1,z_2^{-1}).
\end{equation}

The neutral part $I^{(0)}$ is
\begin{equation}
I^{(0)}=1+x(\chi_1(z_1)+2)
+x^2(2\chi_2(z_1)+2\chi_1(z_1)+3)
+\cdots.
\end{equation}
The non-trivial contributions to the
positive part $I^{(+)}$ up to $x^2$ are
\begin{eqnarray}
I^{(+)}_{(\Y(1),\Y(1))}
&=&
\left[
x\chi_\frac{1}{2}(z_1)
+x^2\chi_\frac{3}{2}(z_1)
+\cdots\right]z_2,\\
I^{(+)}_{(\Y(2),\Y(2))}&=&
x^2\chi_1(z_1)z_2^2+\cdots,\\
I^{(+)}_{(\Y(1,1),\Y(1,1))}&=&x^2\chi_1(z_1)z_2^2+\cdots.
\end{eqnarray}
Combining these contributions, we obtain the complete index $I$ and the corresponding
single-particle index $I^{\rm sp}$,
\begin{eqnarray}
I&=&1+x(\chi^{SU(3)}_{(1,1)}(z_1,z_2)+1)
+x^2(2\chi^{SU(3)}_{(2,2)}(z_1,z_2)+\chi^{SU(3)}_{(1,1)}(z_1,z_2)+1)
+\cdots
,\\
I^{\rm sp}
&=&x(\chi^{SU(3)}_{(1,1)}(z_1,z_2)+1)
+x^2(\chi^{SU(3)}_{(2,2)}(z_1,z_2)-\chi^{SU(3)}_{(1,1)}(z_1,z_2)-1)
+\cdots,
\end{eqnarray}
where
$\chi^{SU(3)}_{(a_1,a_2)}$ is the $SU(3)$ character for the representation
with Dynkin label $(a_1,a_2)$.
We here define $\chi_{(a_1,a_2)}^{SU(3)}$
in a different way from (\ref{su3ch}).
The characters for fundamental and the anti-fundamental representations
are
\begin{equation}
\chi_{(1,0)}^{SU(3)}(z_1,z_2)=z_2^{1/3}\chi_\frac{1}{2}(z_1)+z_2^{-2/3},\quad
\chi_{(0,1)}^{SU(3)}(z_1,z_2)=z_2^{-1/3}\chi_\frac{1}{2}(z_1)+z_2^{2/3}.
\end{equation}

The form of the index strongly suggest the enhancement of the global symmetry to
$SU(3)$.

\section{Conclusions}\label{discussions.sec}
We derived a formula of the superconformal index
for ${\cal N}=2$ superconformal
large $N$ quiver Chern-Simons theories.
In the case of vector-like models,
we confirmed the factorization of the index into
the neutral, the positive, and the negative parts.
For chiral theories, the factorization was confirmed
only in the sector of diagonal monopole operators.
We did not take account of non-diagonal monopole operators
on the assumption that they are non-BPS and do not contribute to the index.
In this paper, we have not proved this on the gauge theory side.
It would be important to analyze the
contribution of non-diagonal operators systematically
for understanding of the relation between non-diagonal
monopole operators and wrapped M2-branes in AdS$_4$/CFT$_3$.

We applied the formula to some Chern-Simons theories
which have been proposed as dual theories to
M-theory in homogeneous spaces, $V^{5,2}$,
$Q^{1,1,1}$,
$Q^{2,2,2}$, $M^{1,1,1}$, and $N^{0,1,0}$.
The actions of these theories do not have full global symmetries
which exist on the gravity side as isometries of the internal spaces.
We computed the index for each theory up to the order of
$x^2$, and confirmed that it is
invariant under the Weyl group of the full symmetry
expected from $AdS/CFT$
if we include the contribution of diagonal monopole operators.
In other wards, the index is a linear combination of characters
of the full symmetry.
This result strongly suggests that the global symmetry
of the Chern-Simons theory is enhanced to the full symmetry
non-perturbatively.

We emphasize that we only confirm the invariance of the index under
the Weyl group, and it does not guarantee
the enhancement of the continuous symmetry.
To obtain more convincing evidences of the symmetry enhancement,
we need to compare the index to the result of
Kaluza-Klein analysis on the gravity side.
Such an analysis is important especially for
inhomogeneous Sasaki-Einstein spaces in which
we have no symmetry enhancement in general.

\section*{Acknowledgements}
Y.I. was supported in part by
Grant-in-Aid for Young Scientists (B) (\#19740122) from the Japan
Ministry of Education, Culture, Sports,
Science and Technology.
D.Y. acknowledges the financial support from the Global
Center of Excellence Program by MEXT, Japan through the
"Nanoscience and Quantum Physics" Project of the Tokyo
Institute of Technology.
S.Y. was supported by
the Global COE Program ``the Physical Sciences Frontier'', MEXT, Japan.

\end{document}